\documentclass{JAC2000}
\usepackage{graphicx}
\newcommand{\ud}{\mathrm{d}}
\begin{document}
\title{NONLINEAR BEAM DYNAMICS AND EFFECTS OF WIGGLERS}
\author{A. Fedorova,  M. Zeitlin, IPME, RAS, V.O. Bolshoj pr., 61,199178, 
St.~Petersburg, Russia\thanks{e-mail: zeitlin@math.ipme.ru}
\thanks{ http://www.ipme.ru/zeitlin.html;
http://www.ipme.nw.ru/zeitlin.html}}

\maketitle

\begin{abstract}
We present the applications of variational--wavelet approach  for
the analytical/numerical treatment of the effects of 
insertion devices on beam
dynamics.
We investigate the dynamical models
which have polynomial nonlinearities and variable coefficients.
We construct the corresponding wavelet representation
for wigglers and undulator magnets.
\end{abstract}

\section{INTRODUCTION}

In this paper we consider the applications of a new nu\-me\-ri\-cal\--analytical 
technique which is based on the methods of local nonlinear harmonic
analysis or wavelet analysis to the treatment of
effects of insertion devices on beam dynamics.      
Our approach in this paper is based on the generalization of variational-wavelet 
approach from [1]-[8],
which allows us to consider not only polynomial but rational type of 
nonlinearities [9]. 
We present solution via
full multiresolution expansion in all time 
scales, which
gives us expansion into a slow part
and fast oscillating parts. So, we may move
from coarse scales of resolution to the 
finest one for obtaining more detailed information about our dynamical process.
In this way we give contribution to our full solution
from each scale of resolution or each time scale.
The same is correct for the contribution to power spectral density
(energy spectrum): we can take into account contributions from each
level/scale of resolution.
Starting from formulation of
initial dynamical problems (part 2) we
construct in part 3 via multiresolution analysis
explicit representation for all dynamical variables in the base of
compactly supported wavelets. 
Then in part 4  we consider further extension of our
previous results to the case of variable coefficients.

\section{Effects of Insertion Devices on Beam Dynamics}

Assuming a sinusoidal field variation, we may consider according to [10]
the analytical treatment of the effects of insertion devices on beam dynamics.
One of the major detrimental aspects of the installation of insertion devices
is the resulting reduction of dynamic aperture. Introduction of non-linearities
leads to enhancement of  the amplitude-dependent tune shifts and distortion of
phase space. The nonlinear fields will produce significant effects at large
betatron amplitudes such as excitation of n--order resonances.
The components of the insertion device vector potential used for the derivation of
equations of motion are as follows:
\begin{eqnarray}
A_x&=&\cosh(k_xx)\cosh(k_yy)\sin(ks)/(k\rho)\\
A_y&=&k_x\sinh(k_xx)\sinh(k_yy)\sin(ks)/(k_yk\rho) \nonumber
\end{eqnarray}
with $k_x^2+k_y^2=k^2=(2\pi/\lambda)^2$,
where $\lambda$ is  the period length of the insertion device, $\rho$ is 
the radius of the curvature in the field $B_0$.
After a canonical transformation to
betatron variables, the Hamiltonian is averaged over the period of
the insertion device and hyperbolic functions are expanded to the fourth
order in $x$ and $y$ (or arbitrary order).
Then we have the following Hamiltonian:
\begin{eqnarray}
H&=&\frac{1}{2}[p_x^2+p_y^2]+\frac{1}{4k^2\rho^2}[k_x^2x^2+k_y^2y^2]\nonumber\\
 &+&\frac{1}{12k^2\rho^2}[k_x^4x^4+k_y^4y^4+3k_x^2k^2x^2y^2]\\
 &-&\frac{\sin(ks)}{2k\rho}[p_x(k_x^2x^2+k_y^2y^2)-2k_x^2p_yxy]\nonumber
\end{eqnarray}
We have in this case also nonlinear (polynomial with degree 3) dynamical system 
with variable (periodic) coefficients. 
After averaging the motion over a magnetic period we
have the following related equations
\begin{eqnarray}
\ddot{x}&=&-\frac{k_x^2}{2k^2\rho^2}\Big[x+\frac{2}{3}k^2_xx^3\Big]-
            \frac{k^2_xxy^2}{2\rho^2}\\
\ddot{y}&=&-\frac{k_y^2}{2k^2\rho^2}\Big[y+\frac{2}{3}k^2_yy^3\Big]-
            \frac{k^2_xx^2y}{2\rho^2}\nonumber
\end{eqnarray}

\section{Wavelet Framework}

The first main part of our consideration is some variational approach
to this problem, which reduces initial problem to the problem of
solution of functional equations at the first stage and some
algebraical problems at the second stage.
Multiresolution expansion is the second main part of our construction.
Because affine
group of translation and dilations is inside the approach, this
method resembles the action of a microscope. We have contribution to
final result from each scale of resolution from the whole
infinite scale of increasing closed subspaces $V_j$:
$\quad
...V_{-2}\subset V_{-1}\subset V_0\subset V_{1}\subset V_{2}\subset ...
$.
The solution is parameterized by solutions of two reduced algebraical
problems, one is nonlinear and the second are some linear
problems, which are obtained by the method of Connection
Coefficients (CC)[11].
We use compactly supported wavelet basis.
Let our  wavelet
expansion be
\begin{eqnarray}
f(x)=\sum\limits_{\ell\in{\bf Z}}c_\ell\varphi_\ell(x)+
\sum\limits_{j=0}^\infty\sum\limits_{k\in{\bf
Z}}c_{jk}\psi_{jk}(x)
\end{eqnarray}
If $c_{jk}=0$ for $j\geq J$, then $f(x)$ has an alternative
expansion in terms of dilated scaling functions only
$
f(x)=\sum\limits_{\ell\in {\bf Z}}c_{J\ell}\varphi_{J\ell}(x)
$.
This is a finite wavelet expansion, it can be written solely in
terms of translated scaling functions.
To solve our second associated linear problem we need to
evaluate derivatives of $f(x)$ in terms of $\varphi(x)$.
Let be $
\varphi^n_\ell=\ud^n\varphi_\ell(x)/\ud x^n
$.
We consider computation of the wavelet - Galerkin integrals.
Let $f^d(x)$ be d-derivative of function
 $f(x)$, then we have
$
f^d(x)=\sum_\ell c_l\varphi_\ell^d(x)
$,
and values $\varphi_\ell^d(x)$ can be expanded in terms of
$\varphi(x)$
\begin{eqnarray}
\varphi_\ell^d(x)&=&\sum\limits_m\lambda_m\varphi_m(x),\\
\lambda_m&=&\int\limits_{-\infty}^{\infty}\varphi_\ell^d(x)\varphi_m(x)\ud x,\nonumber
 \end{eqnarray}
where $\lambda_m$ are wavelet-Galerkin integrals.
The coefficients $\lambda_m$  are 2-term connection
coefficients. In general we need to find $(d_i\geq 0)$
\begin{eqnarray}
\Lambda^{d_1 d_2 ...d_n}_{\ell_1 \ell_2 ...\ell_n}=
 \int\limits_{-\infty}^{\infty}\prod\varphi^{d_i}_{\ell_i}(x)dx
\end{eqnarray}
For Riccati case we need to evaluate two and three
connection coefficients
\begin{eqnarray}
&&\Lambda_\ell^{d_1
d_2}=\int^\infty_{-\infty}\varphi^{d_1}(x)\varphi_\ell^{d_2}(x)dx,
\\
&&\Lambda^{d_1 d_2
d_3}=\int\limits_{-\infty}^\infty\varphi^{d_1}(x)\varphi_
\ell^{d_2}(x)\varphi^{d_3}_m(x)dx \nonumber
\end{eqnarray}
According to CC method [11] we use the next construction. When $N$  in
scaling equation is a finite even positive integer the function
$\varphi(x)$  has compact support contained in $[0,N-1]$.
For a fixed triple $(d_1,d_2,d_3)$ only some  $\Lambda_{\ell
 m}^{d_1 d_2 d_3}$ are nonzero: $2-N\leq \ell\leq N-2,\quad
2-N\leq m\leq N-2,\quad |\ell-m|\leq N-2$. There are
$M=3N^2-9N+7$ such pairs $(\ell,m)$. Let $\Lambda^{d_1 d_2 d_3}$
be an M-vector, whose components are numbers $\Lambda^{d_1 d_2
d_3}_{\ell m}$. Then we have the first reduced algebraical system
: $\Lambda$
satisfy the system of equations $(d=d_1+d_2+d_3)$
\begin{eqnarray}
&&A\Lambda^{d_1 d_2 d_3}=2^{1-d}\Lambda^{d_1 d_2 d_3},
\\
&&A_{\ell,m;q,r}=\sum\limits_p a_p a_{q-2\ell+p}a_{r-2m+p}\nonumber
\end{eqnarray}
By moment equations we have created a system of $M+d+1$
equations in $M$ unknowns. It has rank $M$ and we can obtain
unique solution by combination of LU decomposition and QR
algorithm.
The second  reduced algebraical system gives us the 2-term connection
coefficients ($d=d_1+d_2$):
\begin{eqnarray}
A\Lambda^{d_1 d_2}=2^{1-d}\Lambda^{d_1 d_2},\quad 
A_{\ell,q}=\sum\limits_p a_p a_{q-2\ell+p}
\end{eqnarray}
For nonquadratic case we have analogously additional linear problems for
objects (6).
Solving these linear problems we obtain the coefficients of reduced nonlinear
algebraical system and after that we obtain the coefficients of wavelet
expansion (4). As a result we obtained the explicit time solution  of our
problem in the base of compactly supported wavelets.
On Fig.1 we present an example of basis wavelet function which satisfies
some boundary conditions. 
In the following we consider extension of this approach to the case of arbitrary 
variable coefficients.

\section{Variable Coefficients}

In the case when we have the situation when our problems (2),(3)
are described by a system of
nonlinear (rational) differential equations, we need to consider
also the extension of our previous approach which can take into  account
any type of variable coefficients (periodic, regular or singular).
We can produce such approach if we add in our construction additional
refinement equation, which encoded all information about variable
coefficients [12].
According to our variational approach we need to compute only additional
integrals of
the form
\begin{equation}\label{eq:var1}
\int_Db_{ij}(t)(\varphi_1)^{d_1}(2^m t-k_1)(\varphi_2)^{d_2}
(2^m t-k_2)\ud x,
\end{equation}
where  $b_{ij}(t)$ are arbitrary functions of time and trial
functions $\varphi_1,\varphi_2$ satisfy the refinement equations:
\begin{equation}
\varphi_i(t)=\sum_{k\in{\bf Z}}a_{ik}\varphi_i(2t-k)
\end{equation}
If we consider all computations in the class of compactly supported wavelets
then only a finite number of coefficients do not vanish. To approximate
the non-constant coefficients, we need choose a different refinable function
$\varphi_3$ along with some local approximation scheme
\begin{equation}
(B_\ell f)(x):=\sum_{\alpha\in{\bf Z}}F_{\ell,k}(f)\varphi_3(2^\ell t-k),
\end{equation}
where $F_{\ell,k}$ are suitable functionals supported in a small neighborhood
of $2^{-\ell}k$ and then replace $b_{ij}$ in (\ref{eq:var1}) by
$B_\ell b_{ij}(t)$. In particular case one can take a characteristic function
and can thus approximate non-smooth coefficients locally. To guarantee
sufficient accuracy of the resulting approximation to (\ref{eq:var1})
it is important to have the flexibility of choosing $\varphi_3$ different
from $\varphi_1, \varphi_2$. In the case when D is some domain, we
can write
\begin{equation}
b_{ij}(t)\mid_D=\sum_{0\leq k\leq 2^\ell}b_{ij}(t)\chi_D(2^\ell t-k),
\end{equation}
where $\chi_D$ is characteristic function of D. So, if we take
$\varphi_4=\chi_D$, which is again a refinable function, then the problem of
computation of (\ref{eq:var1}) is reduced to the problem of calculation of
integral
\begin{eqnarray}
&&H(k_1,k_2,k_3,k_4)=H(k)=
\int_{{\bf R}^s}\varphi_4(2^j t-k_1)\cdot \nonumber\\
&&\varphi_3(2^\ell t-k_2)
\varphi_1^{d_1}(2^r t-k_3)
\varphi_2^{d_2}(2^st-k_4)\ud x
\end{eqnarray}
The key point is that these integrals also satisfy some sort of refinement
equation [12]:
\begin{equation}
2^{-|\mu|}H(k)=\sum_{\ell\in{\bf Z}}b_{2k-\ell}H(\ell),\qquad \mu=d_1+d_2.
\end{equation}
This equation can be interpreted as the problem of computing an eigenvector.
Thus, we reduced the problem of extension of our method to the case of
variable coefficients to the same standard algebraical problem as in
the preceding sections. So, the general scheme is the same one and we
have only one more additional
linear algebraic problem by which we can parameterize the
solutions of corresponding problem in the same way.

On Fig.~2 we present approximated configuration and on Fig.~3 the 
corresponding
multiresolution representation according to formula (4).
\begin{figure} [htb] 
\centering
\includegraphics*[width=60mm]{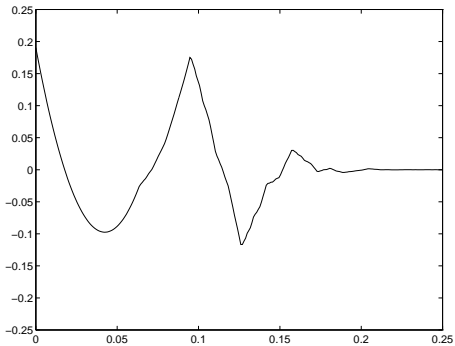}
\caption{Basis wavelet with fixed boundary conditions}
\end{figure}
\begin{figure} [htb] 
\centering
\includegraphics*[width=60mm]{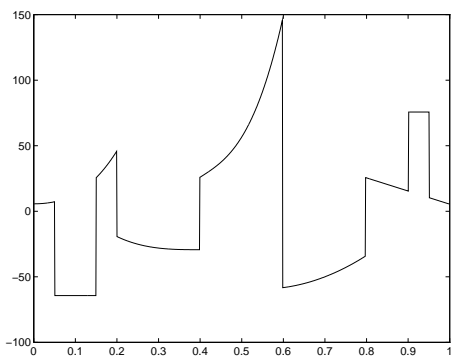}
\caption{Approximated configuration}
\end{figure}
\begin{figure} [t]
\centering
\includegraphics*[width=60mm]{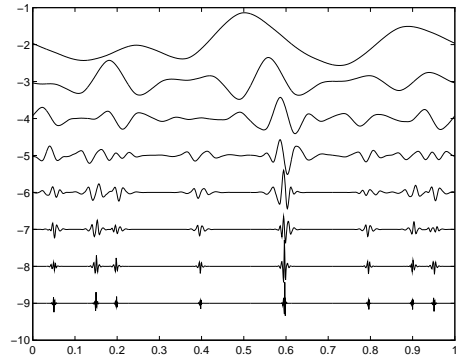}
\caption{Multiresolution representation}
\end{figure}

We would like to thank Professor
James B. Rosenzweig and Mrs. Melinda Laraneta for
nice hospitality, help and support during 
UCLA ICFA Workshop.

\end{document}